\documentstyle[11pt,epsfig]{article}

\begin{document}

\begin{titlepage}
\begin{center}
\large STATE RESEARCH CENTER OF RUSSIA\\
INSTITUTE FOR HIGH ENERGY PHYSICS\\
\end{center}

\bigskip

\begin{flushright}
{ IHEP 97-67\\}
\end{flushright}     

\bigskip

\begin{center}{\Large The search for sleptons and
flavour lepton number violation at LHC (CMS)}
\end{center}

\bigskip

  \begin{center}
{\large
    S.I.~Bityukov (IHEP, Protvino RU-142284, Russia),\\
    N.V.~Krasnikov (INR, Moscow 117312, Russia)
}
\end {center}

\vspace{2cm}
  
\begin{abstract}

We study a possibility to detect sleptons and flavour lepton number
violation at LHC (CMS). We investigate the production and decays 
of right- and left-handed sleptons separately. We have found that for
$L~=~10^5pb^{-1}$ it would be possible to discover right-handed sleptons
with a mass up to 350~GeV and left-handed ones with a mass up to 350~GeV. 
We also investigate a possibility to look for flavour lepton number
violation in slepton decays due to the mixing of different generations
sleptons. We find that for the maximal $(\tilde \mu_{R}-\tilde e_{R})$
mixing it is possible to detect such effect for sleptons with a mass
up to 250~GeV.
\end{abstract}

\vspace{4cm}

\begin{center}
Protvino 1997
\end{center}

\end{titlepage}

\newpage

\section{Introduction}

As is well known, one of the supergoals of LHC is the supersymmetry
discovery. In particular, it is very important to investigate a possibility
to discover nonstrongly interacting superparticles (sleptons,
higgsino, gaugino). In this paper we investigate the discovery potential
for sleptons and for flavour lepton number violation in slepton decays at CMS.

This study is distinct from the previous ones~\cite{1,2,3} in several aspects.
We do not use the minimal SUGRA-MSSM framework. Instead we investigate 
separately the production and decays of right-handed and left-handed sleptons.
We believe that SUGRA-MSSM framework is very particular and very attractive
model. However, the main assumption of SUGRA-MSSM model is that at GUT scale
($M_{GUT}=2 \cdot 10^{16}~GeV$) all the soft breaking scalar masses coincide
could be, at best, a rough feature due to many reasons:

\begin{enumerate}
\item
in superstring inspired models soft scalar supersymmetry breaking terms
are not universal at Planck scale in general~\cite{4},
\item
in supersymmetric SU(5) model an account of the evolution of soft 
supersymmetry breaking terms between Planck and GUT scale~\cite{5,6} 
is very essential,
\item 
in models with additional relatively light vector like supermultiplet
the mass formulae for superparticles can drastically differ~\cite{7}
from the standard ones~\cite{8}.
\end{enumerate}

\noindent
Therefore, we believe that it is more appropriate not to rely on the 
particular model but try to investigate model independent aspects of 
the sleptons search at LHC. The cross section for the production of the
right(left)-handed sleptons depends mainly on the mass of right(left)-handed
sleptons and the decay properties of the sleptons are determined mainly 
by the mass of the lightest superparticle (LSP).

The main signature for the search for sleptons at LHC is two same-flavour
opposite-sign leptons $+ E^{miss}_T + no~jets$. Of course, there are SUSY
strong $(\tilde g \tilde g,~\tilde g \tilde q,~\tilde q \tilde q)$ and 
weak $(\chi^{\pm}_1 \chi^0_2,~\chi^{\pm}_1 \chi^{\mp}_1)$ backgrounds
to the slepton signature. But, as a rule, they are not too large and they
only increase the LHC discovery potential of new SUSY physics
(however, in general, it could be nontrivial to separate a slepton signal
from SUSY backgrounds).

As it has been mentioned above, we study separately
both the right-handed slepton production and the signature and the  
left-handed slepton production and the signature. We also investigate
the possibility to look for the flavour lepton number violation
in slepton decays at LHC. We find that at LHC for $L~=~10^5pb^{-1}$
it is possible to discover right- and left-handed sleptons with a mass
up to 350~GeV and
the flavour number violation in right-handed slepton decays for
slepton masses up to 250 GeV as well.

Our simulations are made at the particle level with parametrized
detector responses based on the detailed detector simulation.
All SUSY processes with full particle spectrum, couplings, 
production cross section and decays are generated with ISAJET~7.13,
ISASUSY~\cite{9}. The Standard Model backgrounds are generated 
with PYIHIA~5.7~\cite{10}. We have used cuts and estimates for the
Standard Model backgrounds obtained in ref.~\cite{3}. The CMS detector
simulation program CMSJET~3.2~\cite{11} is used.

In Section 1 we describe slepton production and decay mechanisms.
Section 2 is devoted to the discussion of the Standard Model backgrounds.
In Sections 3, 4 and 5 we discuss the case of right-handed, 
left-handed and left- plus right-handed sleptons, correspondingly.
Section 6 is devoted to the discussion of the search for flavour lepton 
violation in slepton decays.

\section{Slepton production and decays}

\par
Relatively heavy sleptons with the masses larger than chargino and neutralino
masses $\chi^{\pm}_1,~\chi^0_2$ can be produced at LHC only through
a Drell-Yan mechanism, namely, pairs 
$\tilde l_L~\tilde l_L,~\tilde l_R~\tilde l_R,~\tilde \nu_L~\tilde \nu_L,~
\tilde \nu_L~\tilde l_L$ can be produced. In general, the decays of sleptons
can be rather complicated. In this section  we shall study the case
when only right-handed sleptons are relatively light, whereas the left-handed
sleptons are heavy and their contribution to the signature 
$l^+l^-~+~E^{miss}_T~+~no~jets$ is small. In this case right-handed
sleptons decay dominantly to LSP
\begin{center}
$\tilde l^-_R \longrightarrow l^- + \chi^0_1$.
\end{center}
If decays to the second neutralino or first chargino are kinematically allowed,
the signature $l^+l^-~+~E^{miss}_T~+~no~jets$ can be realized as a result 
of the gaugino decays

\hspace{4cm} $\chi^0_2 \longrightarrow \chi^0_1 + l^+l^-$

\hspace{4cm} $\chi^0_2 \longrightarrow \chi^0_1 + \nu \bar \nu$

\hspace{4cm} $\chi^0_2 \longrightarrow \chi^0_1 + Z$~~~~~~~~~~~~~~~(A)

\hspace{4cm} $\chi^{\pm}_1 \longrightarrow \chi^0_1 + l^{\pm} + \nu$

\hspace{4cm} $\chi^{\pm}_1 \longrightarrow \chi^0_1 + W^{\pm}.$

\noindent
For the case when $\chi^0_2,~\chi^{\pm}_1$ are heavier than sleptons,
an indirect slepton production is possible

\hspace{4cm} $\chi^0_2 \longrightarrow \tilde l^{\pm}_{L,R}l^{\mp}$

\hspace{4cm} $\chi^0_2 \longrightarrow \tilde \nu_L \bar \nu_L$~~~~~~~~~~~~~~~~~(B)

\hspace{4cm} $\chi^{\pm}_1 \longrightarrow \tilde \nu_L l^{\pm}$

\hspace{4cm} $\chi^{\pm}_1 \longrightarrow \tilde e^{\pm}_L \nu_L.$

\noindent
The left-handed sleptons decay (if kinematically accessible) 
to charginos and neutralinos

\hspace{4cm} $\tilde l^{\pm}_L \longrightarrow l^{\pm} + \chi^0_{1,2}$

\hspace{4cm} $\tilde l^{\pm}_L \longrightarrow \nu_L + \chi^{\pm}_1$~~~~~~~~~(C)

\hspace{4cm} $\tilde \nu_L \longrightarrow \nu_L + \chi^0_{1,2}$

\hspace{4cm} $\tilde \nu_L \longrightarrow l^{\pm} + \chi^{\mp}_1.$

In Section 2 we shall discuss the pure left-handed case. In our study we, 
as a rule, neglect indirect slepton production (case~B). Inclusion
of indirect slepton production or other SUSY backgrounds only increases
the excess of the signal over background and improves the significance.
Moreover, for many kinematical points indirect production is not too large.
So, in the first approximation, we neglect cascade decays (case~B) and
(case~C). For the lightest chargino and two lightest neutralinos
in the assumption that at GUT scale 
$M_{GUT} \approx 2 \cdot 10^{16}GeV$ all gaugino masses coincide,
the masses are determined by the common gaugino mass $m_{\frac{1}{2}}$
at GUT scale:

\begin{center}
$m(\chi^0_2) \approx m(\chi^{\pm}_1) \approx 2~m(\chi^0_1) 
\approx m_{1 \over 2}.$
\end{center}

In MSUGRA model slepton masses are determined by formulae~\cite{8}:

\begin{equation}
m^2_{\tilde l_R} = m^2_0 + 0.15~m^2_{1 \over 2} - 
sin^2\theta_WM^2_Zcos2\beta
\end{equation}

\begin{equation}
m^2_{\tilde l_L} = m^2_0 + 0.52~m^2_{1 \over 2} - 
{1 \over 2}(1 - 2~sin^2\theta_W)M^2_Zcos2\beta
\end{equation}

\begin{equation}
m^2_{\tilde \nu} = m^2_0 + 0.52~m^2_{1 \over 2} - 
{1 \over 2}M^2_Zcos2\beta,
\end{equation}

\noindent
where $m_0$ is the common scalar soft breaking mass at GUT scale.
However, formulae (1-3) are valid only within MSUGRA model which 
can be considered, in the best case, as a first rough approximation
(see discussion in the Introduction) and will be wrong for more complicated
models. In this paper we investigate only the signature
$l^+l^-~+~E^{miss}_T~+~no~jets$ which arises as a result of
the slepton pairs production with their subsequent decays into leptons 
and LSP

\begin{center}
$pp \longrightarrow (\tilde l^{\pm} \rightarrow l^{\pm} + ...) +
(\tilde l^{\mp} \rightarrow l^{\mp} + ...).$
\end{center}

\section{Standard Model backgrounds}

The expected main Standard Model background should be $t \bar t$
production, with both W's decaying to leptons, or one of the leptons 
from W decay and the other from the $b-$decay of the same $t-$quark;
the other SM backgrounds come from WW, WZ, $b \bar b$ and 
$\tau \tau$-pair production, with decays to electrons and muons.
Standard model backgrounds for different kinematical cuts have been 
calculated~\cite{3} with the help of PYTHIA~5.7 code. In this paper
we use the results from ref.~\cite{3}. 

The set of kinematical variables which are useful to extract the slepton
signals and typical selection cuts are~\cite{1,2,3}:
\begin{itemize}
\item[i)] for leptons~:

\begin{itemize}
\item
$p_T-$cut on leptons and lepton isolation (Isol), which is here defined 
as the calorimetric energy flow around the lepton in a cone
$\Delta R~<~0.5$ divided by the lepton energy;
\item 
effective mass of two same-flavour opposite-sign leptons, to suppress 
$WZ$ and potential $ZZ$ backgrounds by rejecting events in
a $m_Z \pm \delta m_Z$ band;
\item
$\Delta \Phi(l^+l^-)-$relative azimuthal angle between two same-flavour
opposite-sign leptons;
\end{itemize}

\item[ii)] for $E^{miss}_T~:$

\begin{itemize}
\item
$E^{miss}_T-$cut,
\item
$\Delta \Phi(E^{miss}_T,ll)-$relative azimuthal angle between 
$E^{miss}_T$ and the resulting dilepton momentum in the transverse plane;
\end{itemize}

\item[iii)] for jets~:

\begin{itemize}
\item
"jet veto"-cut~: $N_{jet}~=~0$ for some $E^{jet}_T$ threshold,
in some rapidity interval, typically $|\eta_{jet}|~<~4.5.$
\end{itemize}

\end{itemize}

Namely, we adopt from the ref.~\cite{3} the set of cuts
which in our notations look as follow~:

\begin{center}
\begin{tabular}{|l|l|l|l|l|l|l|l|}
\hline
Cut~~$\backslash$~~Set & 1 & 2 & 3 & 4 & 5 & 6 & 7 \\
\hline
$P^l_L~>$  & 20~GeV & 20~GeV &50 GeV&50 GeV& 60 GeV& 60 GeV& 60 GeV\\
$Isol~<$   &    0.1 &  0.1   & 0.1 &0.1&0.1&0.1&0.03\\
$|\eta_l|~<$             &    2.5 &  2.5   & 2.5 &2.5&2.5&2.5&2.5\\
\hline
$E^{miss}_T~>$  & 50~GeV & 50~GeV & 100 GeV&120 GeV&150 GeV&150 GeV&150 GeV\\
\hline
$\Delta \Phi(E^{miss}_T,ll)>$ &
$160^o$& $160^o$  &$150^o$&$150^o$&$150^o$&$150^o$&$150^o$\\
\hline
$N_{jet}~=$              &      0 &  0     & 0& 0& 0&0&0\\
$E^{jet}_T~>$      & 30 GeV & 30 GeV & 30 GeV&30 GeV&45 GeV&45 GeV&45 GeV\\
$|\eta_{jet}|~<$      &    4.5 &    4.5 & 4.5&4.5&4.5&4.5&4.5\\
\hline
$M_Z-cut$& yes & yes  & yes & yes & yes & yes & yes \\
\hline
$\Delta \Phi(l^+l^-)$  &  
$>130^o$ & no &$<130^o$ &$<130^o$&$<130^o$&$<140^o$&$<130^o$\\
\hline
$N^{SM}_B$~\cite{3}    & 992 & 2421 & 172 & 105 & 45 & 53& 38\\
\hline
\end {tabular}
\end{center} 

\noindent
Here $N^{SM}_B$ is the number of the Standard Model background events 
for $L~=~10^4pb^{-1}$ (cuts 1-2) and for $L~=~10^5pb^{-1}$ (cuts 3-7).
$M_Z-cut$ is condition that $M_{l^+l^-}~<~86~GeV$ or $96~GeV~<~M_{l^+l^-}$.

As has been mentioned above, we, as a rule, neglect indirect
slepton production and also we neglect SUSY backgrounds which are mainly due to
$\tilde q \tilde q,~\tilde g \tilde q,~\tilde g \tilde g$ production and with 
subsequent cascade decays with jets outside the acceptance or below the 
threshold.
As has been demonstrated in ref.~\cite{3} by the example of MSUGRA model
SUSY background is, as a rule, much less than the SM background and we shall
neglect it. At any rate, the SUSY background increases an excess of a signal
over the SM background and increases the LHC discovery potential of new 
physics.
  
\section{Right-handed sleptons}

In this section we study the possibility to search for right-handed sleptons
at CMS. Namely, we consider the signature {\it dilepton +
$E^{miss}_T$~+~no~jets}. We don't consider the left-handed slepton  
contribution to this signature, i.e. we consider the situation when left-handed
sleptons are much heavier than the right ones and it is possible to neglect
them. In this case right-handed sleptons decay dominantly to an LSP

\begin{center}
$\tilde l^-_R \rightarrow l^- + \chi^0_1$.
\end{center}

\noindent
The cross section of the right-handed slepton production is determined
mainly by the mass of the right-handed slepton. The dependence of the 
right-handed slepton cross section production for the case of 3-flavour 
degenerate right-handed  
sleptons is presented in Table~1 and in Fig.1. 

\begin{table}[h]
    \caption{The cross section $\sigma(p p \rightarrow \tilde L_R \tilde L_R
\, + \, ...)$ in pb for different values of right-handed slepton masses at LHC.
Right-handed sleptons are assumed to be degenerate in mass.}
    \label{tab:Tab.1}
    \begin{center}
\begin{tabular}{|l|l|l|l|l|l|l|l|}
\hline
$M(GeV)$ &    90 &   100  &   125  &   150  &   175  &   200  &   225 \\ 
\hline
$\sigma$ &  0.41 &   0.27 &   0.13 &  0.068 &  0.039 &  0.024 &  0.016\\
\hline
\hline
$M(GeV)$ &  250  &   275  &   300  &   325  &   350  &   375  &   400 \\ 
\hline
$\sigma$ & 0.011 & 0.0079 & 0.0055 & 0.0041 & 0.0032 & 0.0025 & 0.0020 \\
\hline
\end{tabular}
    \end{center}
\end{table}

\noindent
The number of signal events
passing through the cuts 1-7 depends rather strongly on the LSP mass
$m_{\chi^0_1}$. The results of our calculations for different values of
the slepton and the LSP masses are presented in Tables~$3-14$. In Tables 
the significances $S$ and $S_{1.5}$ are

\begin{equation}
S = \frac{N_S}{\sqrt{N_S~+~N^{SM}_B}},~~~ 
S_{1.5} = \frac{N_S}{\sqrt{N_S~+~1.5~\cdot~N^{SM}_B}}, 
\end{equation}

\noindent
where $N^{SM}_B$ have been calculated in ref.~\cite{3}. 
The significance $S_{1.5}$ is the significance for the case 
when the number of the SM background events 
is increased by a factor of 1.5 compared to $N^{SM}_B$ 
calculated in ref.~\cite{3}.

As it follows from Tables $3-14$ for $L~=~10^5pb^{-1}$,
it is possible to discover the right-handed sleptons at 5~$\sigma$ 
significance level with a mass up to 300~GeV. For the right-handed sleptons
with a mass $90~GeV~\leq~m_{\tilde l_R}~\leq~300~GeV$ it is possible
to discover sleptons for not very large values of the LSP mass. Typically
we must have  $m_{\chi^0_1}~\leq~(0.4~-~0.6)m_{\tilde l_R}.$

It should be noted that the SM background coming from 
$W~W,~\bar t~t,~\bar \tau~\tau$ production with subsequent  leptonic decays 
predicts the equal number of $\mu^+~\mu^-,~e^+~e^-,~\mu^+~e^-$ and 
$e^+~\mu^-$ events up to statistical fluctuation whereas the signal 
contains an equal number of $\mu^+~\mu^-$ and $e^+~e^-$ pairs coming from

\begin{center}
$p~p \rightarrow \tilde \mu^+_R~\tilde \mu^-_R~+~\dots~
\rightarrow \mu^+~\mu^-~+~2~LSP$ 
\end{center}

and

\begin{center}
$p~p \rightarrow \tilde e^+_R~\tilde e^-_R~+~\dots~\rightarrow \mu^+~\mu^-~+~
2~LSP.$ 
\end{center}

\noindent
We have found that the reaction 

\begin{center}
$p~p \rightarrow \tilde \tau^+_R~\tilde \tau^-_R~+~\dots~
\rightarrow \mu^+~\mu^-,~e^+~e^-,~e^+~\mu^-,~\mu^+~e^-~+~\dots$
\end{center}

\noindent
practically does not contribute to the number of signal events. So, we can 
neglect it, i.e. practically at LHC using the signature 
{\it dilepton + $E_T^{miss}$ + no jets}, we study the production and decays
of the first 2 generations of sleptons.
Therefore, we have qualitative consequence of the existence
of slepton signal -- the excess of 
$\mu^+~\mu^-$ and $~e^+~e^-$ events over $e^+~\mu^-$ and $\mu^+~e^-$ events.

  This excess is other estimation of $N_s$ with
  the statistical fluctuation at $1~\sigma$ level
  equal to $\sqrt{N_s+2N_b}$. If we combine two
  estimations of $N_s$ and adopt $5\sigma$ criterium for
  new physics discovery, we find that the signal events
  passing cuts 1-7 have to be greater than
  143, 220, 64, 52, 38, 40, 35 events, correspondingly. 
  This is an additional criterium for new physics
  discovery for the case of standard slepton production.
  According to this criterium it is possible to discover right-handed sleptons 
  for a mass up to 350~GeV (see Table~14, $m_{\chi^0_1}$~=~119~GeV).
  If we increase the SM~background by a factor of 1.5 which takes 
  into account some
  uncertainties in the calculations of the SM~background, the number of signal
  events passing through the cuts 1-7 has to be greater than 174, 267,
  77, 62, 44, 47, 40 events, correspondingly, and the right-handed sleptons 
  can be discovered with masses up to 325~GeV. At any rate, this criterium
  serves as an additional check for sleptons discovery.

\begin{figure}[htpb]
  \begin{center}
    \resizebox{7cm}{!}{\includegraphics{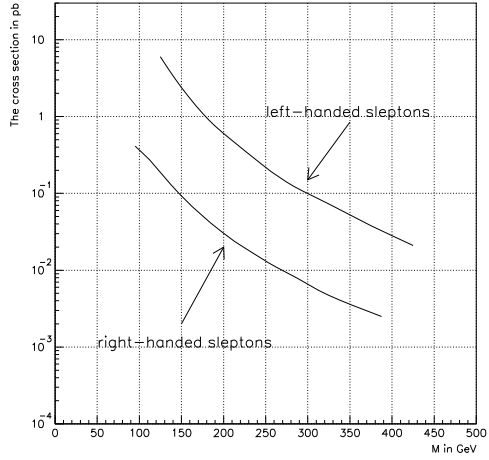}} 
\caption{The cross section $\sigma(p p \rightarrow \tilde L_R \tilde L_R
\, + \, ...)$ in pb for different values of right-handed slepton masses at LHC
(right-handed sleptons are assumed to be degenerate in mass) and
the cross section $\sigma(p p \rightarrow \tilde L_L \tilde L_L
\, + \, ...)$ in pb for different values of left-handed slepton masses 
for $tan \beta~=~2$ at LHC (left-handed sleptons masses are assumed to be 
degenerate in flavour).}
    \label{fig:1} 
  \end{center}
\end{figure}
        
\section{Left-handed sleptons}

In this section we study the case when the right-handed sleptons are much
heavier than the left-handed ones (of course, this situation looks
pathological since in MSUGRA approach the left-handed sleptons are
heavier than the right-handed ones, however, in a general case we can't
exclude such a possibility) and we can neglect them.  The dependence of the 
left-handed slepton cross section production is presented in Table~2 
and in Fig.1. The results of our calculations for different values of the
sleptons and the LSP masses
are presented in Tables 15-20. The notations are similar to the case
of the right-handed sleptons. As follows from Tables 15-20, it is 
possible
to discover left-handed sleptons with a mass up to  350~GeV.  
The discovery potential of the left-handed 
sleptons depends as in the case of the right-handed sleptons on the mass
of LSP. The LSP masses $m_{\chi^0_1}~=~(0.4-0.6)m_{\tilde l_L}$ give the
maximal number of events which passed cuts unlike the case of the right-handed 
sleptons when small LSP masses are the most preferable from the LHC
sleptons discovery point of view. Again, if we use the criterium based
on the estimation of the difference
$N(e^+e^-+\mu^+\mu^-)~-~N(e^+\mu^-+\mu^+e^-)$
events we find that it is possible to discover the left-handed sleptons 
with a mass up to 350~GeV (Table 20, $m_{\chi^0_1}=196~GeV$).

\begin{table}[h]
    \caption{The cross section $\sigma(p p \rightarrow \tilde L_L \tilde L_L
\, + \, ...)$ in pb for different values of left-handed slepton masses 
for $tan \beta =2$ at LHC.
Left-handed sleptons masses are assumed to be degenerate in flavour.}
    \label{tab:Tab.2} 
\begin{center}
\begin{tabular}{|l|l|l|l|l|l|l|l|l|l|}
\hline
$M(GeV)$  & 100 & 150 &  200 &  250 &  300  &  350  &  400  &  450\\
\hline
$\sigma$  & 6.0 & 1.1 & 0.36 & 0.14 & 0.073 & 0.038 & 0.021 & 0.013\\
\hline 
\end{tabular}
\end{center}
 \end{table}

\section{Right-handed plus left-handed sleptons}

As has been mentioned in the Introduction, in general, we can expect
that MSUGRA model, in the best case, gives qualitative description
of the sparticle spectrum. So, in general masses of the LSP, the right-handed 
and the left-handed sleptons are arbitrary. However, in many models the 
left-handed sleptons are, as a rule, heavier than the right-handed ones. 
Moreover,
with a good accuracy the right-handed and the left-handed sleptons give 
an additive contribution to the signal event, i.e.  

\begin{center}
$N(signal)~=~N_{Left}(signal)~+~N_{Right}(signal).$
\end{center}

To obtain some flavour we have studied the direct sleptons production
for the case when $m_{\tilde l_L}~=~m_{\tilde l_R}~+~50GeV$, $tan \beta~=~2.$
The results of our investigation are presented in Tables 21-24.
As follows from Tables 21-24, it is possible to discover
sleptons with a mass of the right-handed slepton up to 350GeV (see Table 24).
The inclusion of the left-handed sleptons increases the sleptons
discovery potential, moreover, it is possible to discover sleptons in 
a wider range of LSP masses. Again the criterium based on the difference 
between
$(e^+e^-+\mu^+\mu^-)$ and $(e^+\mu^-+\mu^+e^-)$ events gives additional 
information about the existence of new physics related with slepton production.

\section{The search for flavour lepton number violation in slepton decays}

As has been mentioned above in MSUGRA the scalar soft supersymmetry 
breaking terms are postulated to be universal at GUT scale. For such
"standard" supersymmetry breaking terms the lepton flavour number is
conserved in supersymmetric extension of the Weinberg-Salam model.
However, in general, squark and slepton supersymmetry breaking mass terms 
are not diagonal due to many reasons~\cite{12} (an account of stringlike
or GUT interactions, nontrivial hidden sector ...) and flavour lepton number
is explicitly broken due to nondiagonal structure of slepton soft 
supersymmetry breaking mass terms. As a consequence such models predict
flavour lepton number violation in $\mu-$ and $\tau-$ decays~\cite{12}.
In ref.~\cite{13,14,15} it has been proposed to look for flavour lepton
number violation in slepton decays at LEP2 and NLC. In ref.~\cite{16}
the possibility to look for flavour lepton number violation in slepton decays 
at LHC has been
studied with 2 points of ref.~\cite{3}. It has been shown
that LHC will be able to discover flavour lepton number violation
in slepton decays for the case of maximal mixing.

In this paper we investigate this problem in a more careful way.
To be specific, consider the case of the mixing between the right-handed
selectron and the right-handed smuon. The mass term for the right-handed
$\tilde e_R' $ and $\tilde \mu_R' $ sleptons has the form

\begin{equation}
-\Delta {\cal L}~=~m^2_1 \tilde e^{+\prime}_R \tilde e^{\prime}_R~+~
m^2_2 \tilde \mu^{+\prime}_R \tilde \mu^{\prime}_R~+~
m^2_{12}(\tilde e^{+\prime}_R \tilde \mu^{\prime}_R~+
~\tilde \mu^{+\prime}_R \tilde e^{\prime}_R).
\end{equation}

\noindent
In formula (5) the last term explicitly violates lepton flavour number.
After the diagonalization of the mass term~(5), 
we find that the eigenstates of the mass term (5) are

\begin{equation}
\tilde e_R~=~\tilde e_R' cos(\phi)~+~\tilde \mu_R' sin(\phi),
\end{equation}

\begin{equation}
\tilde \mu_R~=~\tilde \mu_R' cos(\phi)~-~\tilde e_R' sin(\phi),
\end{equation}

\noindent
with the masses

\begin{equation}
M^2_{12}~=~\frac{1}{2} [(m^2_1 + m^2_2) \pm 
(~(m^2_1 - m^2_2)^2 + 4(m^2_{12})^2~)^{\frac{1}{2}}],
\end{equation}

\noindent
which practically coincide for small values of $m^2_1-m^2_2$ and $m^2_{12}.$
Here the mixing angle is determined by the formulae 

\begin{equation}
tan(2 \phi)~=~\frac{2 m^2_{12}}{m^2_1-m^2_2}.
\end{equation}

\noindent
The crucial point is that even for a small mixing parameter $m^2_{12}$
due to the smallness of $(m^2_1-m^2_2)$ the mixing angle $\phi$ is, in general,
not small (at the present state of art, it is impossible to calculate 
the mixing angle $\phi$ reliably). For the most probable case when the
lightest superparticle is the superpartner of the $U(1)$ gauge boson plus some
small mixing with other gaugino and higgsino, the sleptons
$\tilde \mu_R$, $\tilde e_R$ decay mainly into leptons $\mu$ and $e$ plus
$U(1)$ gaugino $\lambda$. The corresponding terms in the Lagrangian 
responsible for slepton decays are

\begin{equation}
L_1~=~\frac{2 g_1}{\sqrt{2}}
(\bar{e}_R \lambda_L \tilde e_R'~+~\bar{\mu}_R \lambda_L \tilde \mu_R'~+~h.c.),
\end{equation}

\noindent
where $g^2_1 \approx 0.13.$ For the case when mixing is absent the decay width
of the right-handed slepton into lepton plus LSP is given by the formulae

\begin{equation}
\Gamma = \frac{g^2_1}{8 \pi} M_{sl} \Delta_f \approx 
5 \cdot 10^{-3} M_{sl}, 
\end{equation}

\begin{equation}
\Delta_f = (1 - \frac{M^2_{LSP}}{M^2_{sl}})^2,
\end{equation}

\noindent
where $M_{sl}$ and $M_{LSP}$ are the masses of the slepton and the lightest
superparticle ($U(1)-$gaugino) respectively. For the case of nonzero
mixing the Lagrangian (10) in terms of the slepton eigenstates reads

\begin{equation}
L_1~=~\frac{2 g_1}{\sqrt{2}}
[\bar{e}_R \lambda_L (\tilde e_R cos(\phi)~-~\tilde{\mu}_R sin(\phi)) +
\bar{\mu}_R \lambda_L (\tilde \mu_R cos(\phi)~-~\tilde{e}_
R sin(\phi)) ~+~h.c.).
\end{equation}

\noindent
Due to nonzero slepton mixing $(sin(\phi) \neq 0)$ we have lepton flavour
number violation in slepton decays, namely:

\begin{equation}
\Gamma(\tilde \mu_R \rightarrow \mu + LSP)~=~\Gamma cos^2(\phi),
\end{equation}

\begin{equation}
\Gamma(\tilde \mu_R \rightarrow e + LSP)~=~\Gamma sin^2(\phi),
\end{equation}

\begin{equation}
\Gamma(\tilde e_R \rightarrow e + LSP)~=~\Gamma cos^2(\phi),
\end{equation}

\begin{equation}
\Gamma(\tilde e_R \rightarrow \mu + LSP)~=~\Gamma sin^2(\phi).
\end{equation}

At LHC the right-handed sleptons are produced mainly through 
the Drell-Yan mechanism
which is flavour blind in such a way that 
even for nonzero slepton mixing, the cross section
$\sigma(p~p~\rightarrow~\tilde \mu^{\pm}_R\tilde e^{\mp}_R+...)$ vanishes
and the single manifestation of the flavour lepton number violation are
sleptons decays with violation of lepton flavour number.

To be specific, we consider here the most optimistic case of maximal
slepton mixing $(\phi=\frac{\pi}{2})$ and neglect the effects related to 
destructive interference~\cite{14,15,16}. For the case of maximal 
selectron-smuon mixing, the number of signal events coming from slepton
decay is (up to statistical fluctuations)\footnote{As has been mentioned 
above, the contribution of $\tilde \tau_R-$sleptons into the 
$l^+l^-~+~E^{miss}_T~+~no~jets$ signature is practically zero and we
neglect it.}

\begin{equation}
N_{sig}(e^+e^-)~=~N_{sig}(\mu^+e^-)~=~N_{sig}(\mu^-e^+)~=~
\frac{1}{4}N^{no~mix}_{sig}(e^+e^-~+~\mu^+\mu^-).
\end{equation}

\noindent
Therefore, for the case of maximal $(\tilde \mu-\tilde e)$ slepton mixing
we expect equal number of $e^+e^-,~\mu^+\mu^-,~e^+\mu^-,~\mu^+e^-$ events
with $E^{miss}_T$ and with little jet activity unlike the case of the
mixing absence where signal events are only $e^+e^-$ and $\mu^+\mu^-$ and,
as a consequence, for the case of zero mixing we have an excess of 
$(e^+e^-+\mu^+\mu^-)$ events over $(e^+\mu^-+\mu^+e^-)$ events due to
nonzero signal events. The number of signal
$(e^+e^-+\mu^+\mu^-+e^+\mu^-+\mu^+e^-)$ events for the case of maximal
mixing coincides (up to statistical fluctuations) with the number of
$(e^+e^-+\mu^+\mu^-)$ signal events for the mixing absence.
Therefore, in our estimates we can use the results of our calculations performed
for the case of zero mixing. We compare the number of background events

\begin{center}
$N^{SM}_B(e^+e^-+\mu^+\mu^-+e^+\mu^-+\mu^+e^-)~=~
2N^{SM}_B(e^+e^-+\mu^+\mu^-)$ 
\end{center}

\noindent
with the number of signal events

\begin{center}
$N_{signal}(e^+e^-+\mu^+\mu^-+e^+\mu^-+\mu^+e^-)~=~
N^{no~mix}_{signal}(e^+e^-+\mu^+\mu^-).$ 
\end{center}

\noindent
The significance is determined by the formulae

\begin{equation}
S = \frac{N^{mix}_{signal}}{\sqrt{N^{mix}_{signal}+N^{mix}_B}}  = 
\frac{N_S}{\sqrt{N_S+2~N^{SM}_B}},
\end{equation}

\noindent
where $N_S$ and $N^{SM}_B$ are the numbers of the signal and the 
background events
for the case of zero mixing. We adopt the standard criterium according to
which the sleptons will be discovered provided the significance is
bigger than $S~\geq~5.$ As it follows from formulae~(19), the maximal
$(\tilde \mu-\tilde e)$ mixing will be discovered provided the significance
for the detection of $(e^+e^-+\mu^+\mu^-)$ events for the case of the mixing
absence is larger than 7. We have found that for the case
of the right-handed sleptons the $(\tilde \mu-\tilde e)$ mixing and,
 hence, flavour
lepton number violation can be detected for the slepton masses up to 250~GeV.
For the case of the left-handed sleptons we can also search for the mixing
effects. In this case, the $(\tilde \mu-\tilde e)$ mixing can also be detected
for the slepton mass up to 250~GeV.

With the maximal stau-smuon mixing the corresponding formulae
are similar to those given above for the selectron-smuon mixing. 
In this case we
expect the number of $e^+e^-$ signal events to be twice greater than the number 
of $\mu^+\mu^-$ signal events and twice smaller than the number of the
$e^+e^-+\mu^+\mu^-$ signal events for the case of the mixing absence.
Then the significance is

\begin{equation}
S = \frac{N_S(e^+e^-)+N_S(\mu^+\mu^-)}
{\sqrt{N_S(e^+e^-+\mu^+\mu^-)+N_S(e^+e^-)+N_S(\mu^+\mu^-)}}  = 
\frac{{3 \over 4}N_S}{\sqrt{{3 \over 4}N_S+N_B}},
\end{equation}

\noindent
where $N_S$ and $N_B$ are the numbers of signal and background events for the
case of the mixing absence. Again, if the significance for the case of zero
mixing is larger than 6.6, then the significance~(20) will be larger than 5.

For the case of $(\tilde e - \tilde \tau)$ mixing we don't expect 
$\mu^{\pm}e^{\mp}$ signal events as in the case of the mixing absence.
However, for the case of $(\tilde e - \tilde \tau)$ mixing we expect
the excess of $\mu^+\mu^-$ events over $e^+e^-$ events.

In the Standard Model the difference
$N^{back}(e^+e^-)-N^{back}(\mu^+\mu^-)$ is zero up to statistical
fluctuations. At 1 $\sigma$ level the statistical fluctuation is
$\sqrt{N^{SM}_B}$, where $N^{SM}_B$ is the number of
$e^+e^-+\mu^+\mu^-$ background events, whereas 
$N^{sig}(e^+e^-)-N^{sig}(\mu^+\mu^-)~=~0.25~N^{sig}$ (zero~mixing).
Therefore, it is very difficult to distinguish at the 5 $\sigma$ level between 
the mixing absence case and the $(\tilde \mu-\tilde \tau)$ mixing case.

The case of selectron-stau mixing is similar to that of smuon-stau mixing,
the only difference being the interchange 
$e \rightarrow \mu,~~\mu \rightarrow e.$

For the case of maximal selectron-smuon-stau mixing, we expect the equal
number $e^+e^-,~\mu^+\mu^-,~e^+\mu^-$ and $\mu^+e^-$ signal events.
Therefore, this case is very similar to that of the maximal 
$(\tilde \mu-\tilde e)$ mixing. The sole difference is the number 
of signal events 

\begin{center}
$N_S(e^+e^-+\mu^+\mu^-+e^+\mu^-+e^-\mu^+)~=
~\frac{2}{3}N^{zero~mixing}_S(e^+e^-+\mu^+\mu^-).$
\end{center}

\noindent
So, the significance is
$S~=~\frac{{2 \over 3}N_S}{\sqrt{2N_B+N_S}},$ where $N_S$ and $N_B$
is the number of signal events for the case of zero mixing.
The significance $S$ for the maximal
$(\tilde \mu-\tilde e-\tilde \tau)$ mixing will be greater than 5 provided
the corresponding significance for the case of zero mixing 
is larger than 10. So, at CMS it would be extremely difficult to detect
the $(\tilde \mu-\tilde e-\tilde \tau)$ mixing.

\section{Conclusion}

Let us state the main results of this paper. We have studied separetely
the possibility to detect the right-handed sleptons, the left-handed
sleptons and right- plus left-handed sleptons at CMS.
     
For the right-handed sleptons the number of signal events passing through cuts
depends on the mass of the slepton and the mass of the LSP. We have found that 
for $L_t~=~10^5pb^{-1}$ it would be possible to discover the right-handed
sleptons for a mass up to 300~GeV using the standard significance criterium
$S~=~\frac{N_S}{\sqrt{N_S+N_B}}~\geq~5.$ However, taking into account the fact 
that the SM background has an equal number of
$(e^+e^-+\mu^+\mu^-)$ and $(e^+\mu^-+\mu^+e^-)$ events and the signal
contributes only to $(e^+e^-+\mu^+\mu^-)$ events, we can compare the difference
$\Delta N~=~N(e^+e^-+\mu^+\mu^-)~-~N(e^+\mu^-+\mu^+e^-).$ Only signal
events contribute to $\Delta N.$ In the Standard Model $\Delta N$ 
is equal to zero up to statistical fluctuations. Requiring that 
$\Delta N^{signal}~\geq~5 \Delta N^{fluct}~=~5\sqrt{2N_B}$, 
we have found that it is possible to detect the right-handed sleptons
by the measurement of the $\Delta N$ with a mass up to 350~GeV.
At any rate, nonzero $\Delta N$ is an independent and very important check
for the sleptons discovery at LHC. For the case when only the left-handed
sleptons contribute to signal events, we have found that it is possible
to discover the left-handed sleptons with a mass up to 350~GeV.
Again, the measurement of the difference $\Delta N$ allows one to detect
the sleptons with a mass up to 350~GeV (Table~20, $m_{\chi^0_1}=196~GeV$).

For the right-handed sleptons we have found that the number of signal events
decreases with the increase of the LSP mass and, typically, it is possible 
to detect the right-handed sleptons provided the LSP mass 
$m_{LSP}~\leq~0.4~m_{\tilde l_R}.$

For the left-handed sleptons we have found that the number of the signal
events is maximal for $m_{LSP}~=~(0.4-0.6)~m_{\tilde l_L}.$
For the LSP masses in this interval, the CMS left-handed slepton discovery
potential is the maximal one.  Note that these results are in agreement
with the similar observations of ref.~\cite{3}. 

We have also studied 
the case of flavour lepton number violation in slepton decays. For
the case of maximal $(\tilde \mu_R-\tilde e_R)$ mixing we have found 
that the signature qualitatively differs from the case of zero mixing,
namely, in this case we don't have an excess of 
$\Delta N~=~N(e^+e^-+\mu^+\mu^-)~-~N(e^+\mu^-+\mu^+e^-)$ events unlike
the case of zero mixing where $\Delta N~>~0.$ So, it is possible 
to distinguish zero mixing and maximal mixing. We have found that it is 
possible to detect the maximal $(\tilde \mu_R-\tilde e_R)$ mixing for the 
right-handed sleptons with a mass up to 250~GeV. We also considered 
the cases of $(\tilde \mu-\tilde \tau)$ and 
$(\tilde \mu-\tilde e-\tilde \tau)$ mixings. However, for such mixing at
$L~=~10^5pb^{-1}$ it is not so easy to distinguish the  mixings from
the case of the mixing absence. Our conclusion about the possibility
to detect sleptons with a mass up to 300~GeV is in
qualitative agreement with the similar results of ref.~\cite{3}. However,
in our paper we have studied more general situation (we have not assumed
MSUGRA model, which as it has been explained in the Introduction is, 
at best, only a rough description of  the sparticle spectrum). In fact,
the number of signal events passing the cuts depends mainly on the masses
of the right- and the left-handed sleptons and on the LSP mass. So, those 
3 parameters determine the possibility to detect sleptons at CMS.
As a rule, we neglecte cascade neutralino or chargino decays 
resulting in the dilepton signature. However, we have checked that  
(especially for cuts 3-7) their contribution is generally not very large
and, moreover, an account of such contribution increases the significance. 
The reason why we have neglected gaugino decays is that, in general, the masses
of $\chi^0_2,~\chi^{\pm}_1$ are model dependent (they  are determined from
standard but an "ad hoc" assumption that at GUT scale all gaugino masses 
coincide).

\begin{center}
 {\large \bf Acknowledgments}
\end{center}

\par
We are  indebted to the participants of Daniel Denegri seminar on physics 
simulations for useful discussions. We would like to thank Luly Rurua
for providing us her code of the events selections. 
 
\par
The research described in this publication was made possible in part by 
Award No RP1-187 of the U.S. Civilian Research and Development Foundation for 
the Independent States of the Former Soviet Union(CRDF).

\bigskip

\bigskip

\begin{flushright}
{\it Received October 13, 1997}
\end{flushright}

\newpage

\begin{table}[t]
    \caption{The number of events and significances $S$ and $S_{1.5}$ for the 
case of right-handed sleptons, $m_{\tilde l_R}~=~96~GeV,~L~=~10^{4}pb^{-1}$.}
    \label{tab:Tab.3} 
\begin{center}
\begin{tabular}{|l|l|l|}
\hline
                      & cut 1 & cut 2 \\
\hline
$m_{\chi_1}~=~24~GeV$ & 243   &   463 \\
 $S$                  & 6.9   &   8.6 \\
 $S_{1.5}$            & 5.8   &   7.2 \\
\hline
$m_{\chi_1}~=~38~GeV$ & 89    &   180 \\
 $S$                  & 2.7   &   3.5 \\
 $S_{1.5}$            & 2.2   &   2.9 \\
\hline
$m_{\chi_1}~=~53~GeV$ & 34    &    34 \\
 $S$                  & 2.7   &   3.5 \\
 $S_{1.5}$            & 2.2   &   2.9 \\
\hline
\end{tabular}
\end{center}
\end{table}

\begin{table}[b]
    \caption{The number of events and significances $S$ and $S_{1.5}$ for the 
case of right-handed sleptons, $m_{\tilde l_R}~=~100~GeV,~L~=~10^{4}pb^{-1}$.}
    \label{tab:Tab.4} 
\begin{center}
\begin{tabular}{|l|l|l| }
\hline
                      & cut 1     & cut 2 \\
\hline
$m_{\chi_1}~=24~GeV$ &     195.  &   366.\\
 $S$                  &     5.7   &   6.9 \\
 $S_{1.5}$            &     4.8   &   5.8 \\
\hline
 $m_{\chi_1}~=38~GeV$ &     171. &    316. \\
 $S$                  &     5.0  &     6.0 \\
 $S_{1.5}$            &     4.2  &     5.0 \\
\hline
$ m_{\chi_1}~=53~GeV$ &     120. &    219. \\
 $S$                  &     3.6  &    4.3 \\
 $S_{1.5}$           &     3.0  &    3.5 \\
\hline
 $m_{\chi_1}~=69~GeV$ &      48.  &    79. \\
 $S$                  &      1.5  &    1.6 \\
 $S_{1.5}$            &      1.2  &    1.3 \\
\hline
\end{tabular}
\end{center}
\end{table}

\begin{table}[h]
    \caption{The number of events and significances $S$ and $S_{1.5}$
for the case of right-handed sleptons, $m_{\tilde l_R}~=~125~GeV,$
$L~=~10^{5}pb^{-1}$.}
    \label{tab:Tab.5} 
\begin{center}
\begin{tabular}{|l|l|l|l|l|l|l|l| }
\hline
  & cut 1 & cut 2 & cut 3 & cut 4 & cut 5 & cut 6 & cut 7 \\
\hline
$ m_{\chi_1}~=~26~GeV$ & 1086.& 2091.& 189.& 79. & 12. & 27. & 12. \\
$ S $                  & 10.4 & 12.9 & 9.9 & 5.8 & 1.6 & 3.0 & 1.7 \\
$ S_{1.5}$             &  8.6 & 10.7 & 8.9 & 5.1 & 1.3 & 2.6 & 1.4 \\
\hline
$ m_{\chi_1}~=54~GeV$  & 806. & 1632.& 50. & 19. &  0. & 6.  &  1. \\
$ S $                  &  7.8 & 10.2 & 3.4 & 1.7 & 0.0 & 0.8 & 0.2 \\
$ S_{1.5}$             &  6.4 &  8.4 & 2.8 & 1.4 & 0.0 & 0.6 & 0.1 \\
\hline
$ m_{\chi_1}~=85~GeV$  & 446. & 845. &  0. &  0. &  0. &  0. &  0. \\
$ S $                  &  4.4 &  5.3 & 0.0 & 0.0 & 0.0 & 0.0 &  0.0 \\
$ S_{1.5}$             &  3.6 &  4.4 & 0.0 & 0.0 & 0.0 & 0.0 &  0.0 \\
\hline
\end{tabular}
\end{center}
  \end{table}

\begin{table}[h]
    \caption{The number of events and significances $S$ and $S_{1.5}$
for the case of right-handed sleptons, $m_{\tilde l_R}~=~150~GeV,$
$L~=~10^{5}pb^{-1}$.}
    \label{tab:Tab.6} 
\begin{center}
\begin{tabular}{|l|l|l|l|l|l|l|l| }
\hline
  & cut 1 & cut 2 & cut 3 & cut 4 & cut 5 & cut 6 & cut 7 \\
\hline
$ m_{\chi_1}~=24~GeV$ & 626.&1209.& 185.& 127.&  38.&  60.&  38. \\
$ S $                 & 6.1 & 7.6 & 9.8 & 8.3 & 4.2 & 5.6 &  4.4 \\
$ S_{1.5}$            & 5.0 & 6.2 & 8.8 & 7.5 & 3.7 & 5.1 &  3.9 \\
\hline
$ m_{\chi_1}~=53~GeV$ & 595.&1183.& 149.&  83.&  15.&  23.&  15. \\
$ S $                 & 5.8 & 7.4 & 8.3 & 6.1 & 1.9 & 2.6 &  2.1 \\
$ S_{1.5}$            & 4.8 & 6.1 & 7.4 & 5.4 & 1.7 & 2.3 &  1.8 \\
\hline
$ m_{\chi_1}~=69~GeV$ & 472.& 922.& 23. & 5.  &  0. & 1.  &  0.  \\
$ S $                 &  4.6& 5.8 & 1.6 & 0.5 & 0.0 & 0.1 & 0.0  \\
$ S_{1.5}$            &  3.8& 4.8 & 1.4 & 0.4 & 0.0 & 0.1 &  0.0 \\
\hline
\end{tabular}
\end{center}
  \end{table}

\begin{table}[h]
    \caption{The number of events and significances $S$ and $S_{1.5}$
for the case of right-handed sleptons, $m_{\tilde l_R}~=~175~GeV,$
$L~=~10^{5}pb^{-1}$.}
    \label{tab:Tab.7} 
\begin{center}
\begin{tabular}{|l|l|l|l|l|l|l|l| }
\hline
  & cut 1 & cut 2 & cut 3 & cut 4 & cut 5 & cut 6 & cut 7 \\
\hline
$ m_{\chi_1}~=26~GeV$ & 345.& 679.& 155.& 130.&  73.&  92.&  72. \\
$ S $                 & 3.4 &  4.3& 8.6 & 8.5 & 6.7 & 7.6 & 6.9 \\
$ S_{1.5}$            & 2.8 &  3.5& 7.6 & 7.7 & 6.2 & 7.0 & 6.3 \\
\hline
$ m_{\chi_1}~=54~GeV$ & 289.& 648.& 137.& 108.& 49. &  61.&  49. \\
$ S $                 & 2.9 & 4.1 & 7.8 & 7.4 & 5.1 & 5.7 & 5.3 \\
$ S_{1.5}$            & 2.3 & 3.4 & 6.9 & 6.6 & 4.5 & 5.1 & 4.8 \\
\hline
$ m_{\chi_1}~=85~GeV$ & 317.& 647.&  83.&  47.&   7.&  17.&  7. \\
$ S $                 & 3.1 &  4.1& 5.2 & 3.8 & 1.0 & 2.0 & 1.0 \\
$ S_{1.5}$            & 2.6 &  3.4& 4.5 & 3.3 & 0.8 & 1.7 & 0.9 \\
\hline
\end{tabular}
\end{center}
  \end{table}

\begin{table}[h]
    \caption{The number of events and significances $S$ and $S_{1.5}$
for the case of right-handed sleptons, $m_{\tilde l_R}~=~200~GeV,$
$L~=~10^{5}pb^{-1}$.}
    \label{tab:Tab.8} 
\begin{center}
\begin{tabular}{|l|l|l|l|l|l|l|l| }
\hline
  & cut 1 & cut 2 & cut 3 & cut 4 & cut 5 & cut 6 & cut 7 \\
\hline
$ m_{\chi_1}~=24~GeV$ & 229.& 506.& 170.& 152.& 110.& 128.& 110. \\
$ S $                 & 2.3 & 3.2 & 9.2 & 9.5 & 8.8 & 9.5 & 9.0 \\
$ S_{1.5}$            & 1.9 & 2.6 & 8.2 & 8.6 & 8.3 & 8.9 & 8.5 \\
\hline
$ m_{\chi_1}~=53~GeV$ & 248.& 476.& 117.& 106.& 75. & 96. & 75. \\
$ S $                 & 2.5 & 3.0 & 6.9 & 7.3 & 6.8 & 7.9 & 7.1 \\
$ S_{1.5}$            & 2.0 & 2.5 & 6.0 & 6.5 & 6.3 & 7.2 & 6.5 \\
\hline
$ m_{\chi_1}~=85~GeV$ & 231.& 447.&  90.&  73.&  40.&  55.&  40. \\
$ S $                 & 2.3 & 2.8 & 5.6 & 5.5 & 4.3 & 5.3 & 4.5 \\
$ S_{1.5}$            & 1.9 & 2.3 & 4.8 & 4.8 & 3.9 & 4.7 & 4.1 \\
\hline
$ m_{\chi_1}~=119~GeV$&  81.& 175.&  1. &  0. &  0. &  0. &   0. \\
$ S $                 & 0.8 & 1.1 & 0.1 & 0.0 & 0.0 & 0.0 &  0.0 \\
$ S_{1.5}$            & 0.7 & 0.9 & 0.1 & 0.0 & 0.0 & 0.0 &  0.0 \\
\hline
\end{tabular}
\end{center}
  \end{table}

\begin{table}[h]
    \caption{The number of events and significances $S$ and $S_{1.5}$
for the case of right-handed sleptons, $m_{\tilde l_R}~=~225~GeV,$
$L~=~10^{5}pb^{-1}$.}
    \label{tab:Tab.9} 
\begin{center}
\begin{tabular}{|l|l|l|l|l|l|l|l| }
\hline
  & cut 1 & cut 2 & cut 3 & cut 4 & cut 5 & cut 6 & cut 7 \\
\hline
$ m_{\chi_1}~=26~GeV$ & 150.& 268.&  81.&  76.&  69.&  87.&  69. \\
$ S $                 & 1.5 & 1.7 & 5.1 & 5.6 & 6.5 & 7.4 & 6.7 \\
$ S_{1.5}$            & 1.2 & 1.4 & 4.4 & 5.0 & 5.9 & 6.7 & 6.1 \\
\hline
$ m_{\chi_1}~=54~GeV$ & 140.& 264.&  84.&  82.&  63.&  75.&  62. \\
$ S $                 & 1.4 & 1.7 & 5.3 & 6.0 & 6.1 & 6.6 & 6.2 \\
$ S_{1.5}$            & 1.1 & 1.4 & 4.5 & 5.3 & 5.5 & 6.0 & 5.7 \\
\hline
$ m_{\chi_1}~=85~GeV$ & 156.& 306.&  85.&  78.&  54.&  66.&  53. \\
$ S $                 & 1.6 & 2.0 & 5.3 & 5.8 & 5.4 & 6.1 & 5.6 \\
$ S_{1.5}$            & 1.3 & 1.6 & 4.6 & 5.1 & 4.9 & 5.5 & 5.1 \\
\hline
$ m_{\chi_1}~=119~GeV$& 135.& 277.&  68.&  55.&  28.&  33.&  28. \\
$ S $                 & 1.3 & 1.8 & 4.4 & 4.3 & 3.3 & 3.6 & 3.4 \\
$ S_{1.5}$            & 1.1 & 1.4 & 3.8 & 3.8 & 2.9 & 3.1 & 3.0 \\
\hline
\end{tabular}
\end{center}
  \end{table}

\begin{table}[h]
    \caption{The number of events and significances $S$ and $S_{1.5}$
for the case of right-handed sleptons, $m_{\tilde l_R}~=~250~GeV,$
$L~=~10^{5}pb^{-1}$.}
    \label{tab:Tab.10} 
\begin{center}
\begin{tabular}{|l|l|l|l|l|l|l|l| }
\hline
  & cut 1 & cut 2 & cut 3 & cut 4 & cut 5 & cut 6 & cut 7 \\
\hline
$ m_{\chi_1}~=24~GeV$ & 128.& 237.&  84.&  78.&  81.&  92.&  81. \\
$ S $                 & 1.3 & 1.5 & 5.3 & 5.8 & 7.2 & 7.6 & 7.4 \\
$ S_{1.5}$            & 1.0 & 1.2 & 4.5 & 5.1 & 6.6 & 7.0 & 6.9 \\
\hline
$ m_{\chi_1}~=53~GeV$ & 125.& 232.&  82.&  76.&  78.&  91.&  78. \\
$ S $                 & 1.2 & 1.5 & 5.1 & 5.6 & 7.0 & 7.6 &  7.2 \\
$ S_{1.5}$            & 1.0 & 1.2 & 4.4 & 5.0 & 6.5 & 7.0 &  6.7 \\
\hline
$ m_{\chi_1}~=85~GeV$ & 117.& 220.&  78.&  73.&  68.&  80.&  68. \\
$ S $                 & 1.2 & 1.4 & 4.9 & 5.5 & 6.4 & 6.9 & 6.6 \\
$ S_{1.5}$            & 1.0 & 1.2 & 4.3 & 4.8 & 5.8 & 6.3 & 6.1 \\
\hline
$ m_{\chi_1}~=119~GeV$& 116.& 217.&  66.&  61.&  49.&  56.&  49. \\
$ S $                 & 1.2 & 1.4 & 4.3 & 4.7 & 5.1 & 5.4 &  5.3 \\
$ S_{1.5}$            & 0.9 & 1.1 & 3.7 & 4.1 & 4.5 & 4.8 &  4.8 \\
\hline
$ m_{\chi_1}~=157~GeV$& 94. & 187.& 43. &  31.&  15.&  18.&  15. \\
$ S$                  & 0.9 & 1.2 & 2.9 & 2.7 & 1.9 & 2.1 & 2.1 \\
$ S_{1.5}$            & 0.8 & 1.0 & 2.5 & 2.3 & 1.7 & 1.8 & 1.8 \\
\hline
\end{tabular}
\end{center}
  \end{table}

\begin{table}[h]
    \caption{The number of events and significances $S$ and $S_{1.5}$
for the case of right-handed sleptons, $m_{\tilde l_R}~=~275~GeV,$
$L~=~10^{5}pb^{-1}$.}
    \label{tab:Tab.11} 
\begin{center}
\begin{tabular}{|l|l|l|l|l|l|l|l| }
\hline
  & cut 1 & cut 2 & cut 3 & cut 4 & cut 5 & cut 6 & cut 7 \\
\hline
$ m_{\chi_1}~=25~GeV$ &  73.& 130.&  45.&  44.&  40.&  49.&  40. \\
$ S $                 & 0.7 & 0.8 & 3.1 & 3.6 & 4.3 & 4.9 &  4.5 \\
$ S_{1.5}$            & 0.6 & 0.7 & 2.6 & 3.1 & 3.9 & 4.3 &  4.1 \\
\hline
$ m_{\chi_1}~=54~GeV$ &  68.& 139.&  53.&  52.&  49.&  59.&  49. \\
$ S $                 & 0.7 & 0.9 & 3.5 & 4.2 & 5.1 & 5.6 & 5.3 \\
$ S_{1.5}$            & 0.6 & 0.7 & 3.0 & 3.6 & 4.5 & 5.0 & 4.8 \\
\hline
$ m_{\chi_1}~=85~GeV$ &  60.& 115.&  35.&  32.&  31.&  34.&  31. \\
$ S$                  & 0.6 & 0.7 & 2.4 & 2.7 & 3.6 & 3.6 & 3.7 \\
$ S_{1.5}$            & 0.5 & 0.6 & 2.0 & 2.3 & 3.1 & 3.2 & 3.3 \\
\hline
$ m_{\chi_1}~=119~GeV$&  91.& 169.&  56.& 54. & 47. & 57. &  46. \\
$ S $                 & 0.9 & 1.1 & 3.7 & 4.3 & 4.9 & 5.4 & 5.0 \\
$ S_{1.5}$            & 0.7 & 0.9 & 3.2 & 3.7 & 4.4 & 4.9 & 4.5 \\
\hline
\end{tabular}
\end{center}
  \end{table}

\begin{table}[h]
    \caption{The number of events and significances $S$ and $S_{1.5}$
for the case of right-handed sleptons, $m_{\tilde l_R}~=~300~GeV,$
$L~=~10^{5}pb^{-1}$.}
    \label{tab:Tab.12} 
\begin{center}
\begin{tabular}{|l|l|l|l|l|l|l|l| }
\hline
  & cut 1 & cut 2 & cut 3 & cut 4 & cut 5 & cut 6 & cut 7 \\
\hline
$ m_{\chi_1}~=24~GeV$ &  64.& 134.&  59.&  59.&  56.&  65.&  56. \\
$ S $                 & 0.6 & 0.9 & 3.9 & 4.6 & 5.6 & 6.0 & 5.8 \\
$ S_{1.5}$            & 0.5 & 0.7 & 3.3 & 4.0 & 5.0 & 5.4 & 5.3 \\
\hline
$ m_{\chi_1}~=52~GeV$ &  59.& 117.&  50.&  46.&  45.&  53.&  45. \\
$ S$                  & 0.6 & 0.8 & 3.4 & 3.7 & 4.7 & 5.1 & 4.9 \\
$ S_{1.5}$            & 0.5 & 0.6 & 2.8 & 3.2 & 4.2 & 4.6 & 4.5 \\
\hline
$ m_{\chi_1}~=85~GeV$ &  55.& 118.&  49.&  46.&  45.&  54.&  45. \\
$ S $                 & 0.6 & 0.8 & 3.3 & 3.7 & 4.7 &  5.2&  4.9 \\
$ S_{1.5}$            & 0.5 & 0.6 & 2.8 & 3.2 & 4.2 & 4.7 &  4.5 \\
\hline
$ m_{\chi_1}~=119~GeV$&  56.& 114.&  46.&  44.&  41.&  48.&  41. \\
$ S $                 & 0.6 & 0.7 & 3.1 & 3.6 & 4.4 & 4.8 & 4.6 \\
$ S_{1.5}$            & 0.5 & 0.6 & 2.6 & 3.1 & 3.9 & 4.3 & 4.1 \\
\hline
$ m_{\chi_1}~=157~GeV$&  54.& 112.&  38.&  36.&  33.&  40.&  33. \\
$ S $                 & 0.5 & 0.7 & 2.6 & 3.0 & 3.7 & 4.1 & 3.9 \\
$ S_{1.5}$            & 0.4 & 0.6 & 2.2 & 2.6 & 3.3 & 3.7 & 3.5 \\
\hline
$ m_{\chi_1}~=196~GeV$&  52.& 105.&  29.&  25.&  13.&  16.&  13. \\
$ S  $                & 0.5 & 0.7 & 2.0 & 2.2 & 1.7 & 1.9 &  1.8 \\
$ S_{1.5}$            & 0.4 & 0.6 & 1.7 & 1.9 & 1.4 & 1.6 &  1.6 \\
\hline
\end{tabular}
\end{center}
  \end{table}

\begin{table}[h]
    \caption{The number of events and significances $S$ and $S_{1.5}$
for the case of right-handed sleptons, $m_{\tilde l_R}~=~325~GeV,$
$L~=~10^{5}pb^{-1}$.}
    \label{tab:Tab.13} 
\begin{center}
\begin{tabular}{|l|l|l|l|l|l|l|l| }
\hline
  & cut 1 & cut 2 & cut 3 & cut 4 & cut 5 & cut 6 & cut 7 \\
\hline
$ m_{\chi_1}~=53~GeV$ &  24.&  72.&  35.&  35.&  36.&  45.&  35. \\
$ S  $                & 0.2 & 0.5 & 2.4 & 3.0 & 4.0 & 4.5 &  4.1 \\
$ S_{1.5}$            & 0.2 & 0.4 & 2.0 & 2.5 & 3.5 & 4.0 &  3.6 \\
\hline
$ m_{\chi_1}~=85~GeV$ &  44.&  89.&  39.&  38.&  40.&  51.&  40. \\
$ S  $                & 0.4 & 0.6 & 2.7 & 3.2 & 4.3 & 5.0 & 4.5  \\
$ S_{1.5}$            & 0.4 & 0.5 & 2.3 & 2.7 & 3.9 & 4.5 & 4.1  \\
\hline
$ m_{\chi_1}~=119~GeV$&  32.&  74.&  37.&  36.&  33.&  43.&  33. \\
$ S  $                & 0.3 & 0.5 & 2.6 & 3.0 & 3.7 & 4.4 & 3.9  \\
$ S_{1.5}$            & 0.3 & 0.4 & 2.2 & 2.6 & 3.3 & 3.9 & 3.5   \\
\hline
$ m_{\chi_1}~=157~GeV$&  34.&  76.&  31.&  29.&  28.&  37.&   28. \\
$ S  $                & 0.3 & 0.5 & 2.2 & 2.5 & 3.3 & 3.9 &  3.4  \\
$ S_{1.5}$            & 0.3 & 0.4 & 1.8 & 2.1 & 2.9 & 3.4 &  3.0 \\
\hline
$ m_{\chi_1}~=196~GeV$&  32.&  73.&  28.&  26.&  19.&  26.&   19. \\
$ S $                 & 0.3 & 0.5 & 2.0 & 2.3 & 2.4 & 2.9 & 2.5 \\
$ S_{1.5}$            & 0.3 & 0.4 & 1.7 & 1.9 & 2.0 & 2.5 & 2.2 \\
\hline
$ m_{\chi_1}~=233~GeV$&  30.&  62.&  17.&  13.&   4.&   6.&   4. \\
$ S $                 & 0.3 & 0.4 & 1.2 & 1.2 & 0.6 & 0.8 &  0.6 \\
$ S_{1.5}$            & 0.2 & 0.3 & 1.0 & 1.0 & 0.5 & 0.6 &  0.5 \\
\hline
\end{tabular}
\end{center}
  \end{table}

\begin{table}[h]
    \caption{The number of events and significances $S$ and $S_{1.5}$
for the case of right-handed sleptons, $m_{\tilde l_R}~=~350~GeV,$
$L~=~10^{5}pb^{-1}$.}
    \label{tab:Tab.14} 
\begin{center}
\begin{tabular}{|l|l|l|l|l|l|l|l| }
\hline
  & cut 1 & cut 2 & cut 3 & cut 4 & cut 5 & cut 6 & cut 7 \\
\hline
$ m_{\chi_1}~=53~GeV$ &  33.&  68.&  33.&  33.&  33.&  40.&  33. \\
$ S  $                & 0.3 & 0.4 & 2.3 & 2.8 & 3.7 & 4.1 &  3.9 \\
$ S_{1.5}$            & 0.3 & 0.4 & 1.9 & 2.4 & 3.3 & 3.7 &  3.5 \\
\hline
$ m_{\chi_1}~=119~GeV$& 32. &  69.& 35. & 34. & 35. & 36. &  35. \\
$ S $                 & 0.3 & 0.4 & 2.4 & 2.9 & 3.9 & 3.8 & 4.1 \\
$ S_{1.5}$            & 0.3 & 0.4 & 2.0 & 2.5 & 3.5 & 3.3 & 3.6 \\
\hline
$ m_{\chi_1}~=196~GeV$&  30.&  65.&  31.&  31.&  27.&  27.&  27. \\
$ S$                  & 0.3 & 0.4 & 2.2 & 2.7 & 3.2 & 3.0 & 3.3 \\
$ S_{1.5}$            & 0.2 & 0.3 & 1.8 & 2.3 & 2.8 & 2.6 & 2.9 \\
\hline
$ m_{\chi_1}~=233~GeV$&  27.&  61.&  25.&  23.&  18.&  18.&  18. \\
$ S $                 & 0.3 & 0.4 & 1.8 & 2.0 & 2.3 & 2.1 &  2.4 \\
$ S_{1.5}$            & 0.2 & 0.3 & 1.5 & 1.7 & 1.9 & 1.8 &  2.1 \\
\hline
$ m_{\chi_1}~=270~GeV$&  23.&  56.&  12.&   7.&   3.&   3.&   3. \\
$ S  $                & 0.2 & 0.4 & 0.9 & 0.7 & 0.4 & 0.4 & 0.5 \\
$ S_{1.5}$            & 0.2 & 0.3 & 0.7 & 0.5 & 0.4 & 0.3 & 0.4 \\
\hline
\end{tabular}
\end{center}
  \end{table}

\begin{table}[t]
    \caption{The number of events and significances $S$ and $S_{1.5}$ for the 
case of left-handed sleptons, $m_{\tilde l_L}~=~100~GeV,~~L~=~10^{4}pb^{-1}$.}
    \label{tab:Tab.15} 
\begin{center}
\begin{tabular}{|l|l|l| }
\hline
  & cut 1 & cut 2 \\
\hline
$ m_{\chi_1}~=24~GeV$ & 132.& 546. \\
$ S  $                & 3.9 & 10.0 \\
$ S_{1.5}$            & 3.3 &  8.4 \\
\hline
$ m_{\chi_1}~=38~GeV$ & 122.& 372. \\
$ S  $                & 3.7 & 7.0 \\
$ S_{1.5}$            & 3.0 & 5.9 \\
\hline
$ m_{\chi_1}~=53~GeV$ & 421. & 602. \\
$ S  $                & 11.2 & 10.9 \\
$ S_{1.5}$            & 9.6  & 9.3 \\
\hline
$ m_{\chi_1}~=69~GeV$ & 209. & 291. \\
$ S  $                & 6.0  & 5.6 \\
$ S_{1.5}$            & 5.1  & 4.6 \\
\hline
\end{tabular}
\end{center}
\end{table}

\begin{table}[b]
    \caption{The number of events and significances $S$ and $S_{1.5}$ for the 
case of left-handed sleptons, $m_{\tilde l_L}~=~150~GeV,~~L~=~10^{5}pb^{-1}$.}
    \label{tab:Tab.16} 
\begin{center}
\begin{tabular}{|l|l|l|l|l|l|l|l|}
\hline
  & cut 1 & cut 2 & cut 3 & cut 4 & cut 5 & cut 6 & cut 7 \\
\hline
$ m_{\chi_1}~=24~GeV$ & 682.&1815.& 302.& 131.&  32.&  41.&  32. \\
$ S $                 &  6.6& 11.3& 13.9& 8.5 & 3.6 & 4.2 & 3.8 \\
$ S_{1.5}$            &  5.5&  9.3& 12.8& 7.7 & 3.2 & 3.7 & 3.4 \\
\hline
$ m_{\chi_1}~=53~GeV$ & 663.&1762.& 143.& 111.& 113.& 120.& 113. \\
$ S $                 & 6.4& 10.9& 8.1 & 7.6 & 9.0 & 9.1 & 9.2 \\
$ S_{1.5}$            & 5.3&  9.0& 7.1 & 6.8 & 8.4 & 8.5 & 8.7 \\
\hline
$ m_{\chi_1}~=85~GeV$ & 951.&2029.&  72.&  11.&   2.&  23.&  2. \\
$ S $                 &  9.1& 12.5& 4.6 & 1.0 & 0.3 & 2.6 & 0.3 \\
$ S_{1.5}$            &  7.6& 10.4& 4.0 & 0.8 & 0.2 & 2.3 &  0.3 \\
\hline
$ m_{\chi_1}~=119~GeV$& 273.& 485.&   2.&   1.&   1.&   2.&  1. \\
$ S $                 & 2.7 & 3.1 & 0.2 & 0.1 & 0.1 &  0.3&  0.2 \\
$ S_{1.5}$            & 2.2 & 2.5 & 0.1 & 0.1 & 0.1 &  0.2&  0.1 \\
\hline
\end{tabular}
\end{center}
\end{table}


\begin{thebibliography}{99}

\bibitem{1} F.del Aguila and L.I.Ametller, Phys.Lett. {\bf B261}(1991)325.
\bibitem{2} H.Baer, C.Chen, F.Paige and X.Tata,
Phys.Rev. {\bf D49}(1994)3283.
\bibitem{3} D.Denegri, L.Rurura and N.Stepanov,
{\it Detection of Sleptons in CMS, Mass Reach,}
CMS Note CMS TN/96-059, October 1996.
\bibitem{4} V.S.Kaplunovsky and J.Louis, Phys.Lett. {\bf B306}(1993)269.
\bibitem{5} N.Polonsky and A.Pomarol, Phys.Rev.Lett. {\bf 73}(1994)2292.
\bibitem{6} N.V.Krasnikov and V.V.Popov,
{\it PLANCSUSY~-~new program for SUSY masses calculations:
from Planck scale to our reality,} Preprint INR 976TH/96.
\bibitem{7} C.Kolda and J.March-Russel, Phys.Rev. {\bf D55}(1997)4252.
\bibitem{8}  As a review see, for instance: \\
R.Barbieri, Riv.Nuovo Cimento {\bf 11}(1988)1; A.B.Lahanus and 
D.V.Nanopoulos, Phys.Rep. {\bf 145}(1987)1; \\ 
H.E.Haber and G.L.Lane, Phys.Rep. {\bf 117}(1985)75;  \\ 
H.P.Nilles, Phys.Rep.{\bf 110}(1984)1; \\ 
N.V.Krasnikov and V.A.Matveev, Physics at LHC, Preprint INR 0940/97; \\
 hep-ph/9703204.
\bibitem{9}  H.Baer, F.Paige , S.Protopesku and X.Tata, {\it Simulating 
Supersymmetry with ISAJET 7.0/ISASUSY 1.0,} Florida State University Preprint 
EP-930329(1993).
\bibitem{10}T.Sjostrand, {PYTHIA 5.7 and ISAJET 7.4, Physics and Manual,}
CERN-TH.7112/93.
\bibitem{11} S.Abdullin, A.Khanov and N.Stepanov, {\it CMSJET 3.2,
CMSJET 3.5,} CMS Note CMS TN/94-180.
\bibitem{12} F.Barzumanti and A.Masieno, Phys.Rev.Lett. {\bf 57}(1986)961;\\
G.K.Leontanis, K.Tamvakis and J.D.Vergados, Phys.Lett. {\bf B171}(1986)412;\\
J.Hagelin, S.Kelley and T.Tanaka, Nucl.Phys. {\bf B415}(1994)293;\\
F.Gabriani and A.Masiero, Nucl.Phys. {\bf B322}(1989)235;\\
I.Antoniadis, J.Ellis, J.S.Hagelin and D.V.Nanopoulos,
Phys.Lett. {\bf B231}(1989)65;\\
S.Kelley, J.L.Lopez, D.V.Nanopoulos and H.Pois,
Nucl.Phys. {\bf B358}(1991)27.
\bibitem{13} N.V.Krasnikov, Mod.Phys.Lett {\bf A9}(1994)2825. 
\bibitem{14} N.V.Krasnikov, Phys.Lett {\bf B388}(1996)783.  
\bibitem{15} Nima Arkani-Hamed, Hsin-Chia Cheng, J.L.Feng and L.J.Hall,
Phys.Rev.Lett. {\bf 77}(1996)1937.
\bibitem{16} N.V.Krasnikov, Zhetp.Lett {\bf 65}(1997)139. 

\end{thebibliography}
\end{document}